\documentclass[%amsmath,amssymb,floatfix,
12pt]{iopart} 

\usepackage{amsfonts}
\usepackage{graphicx}% Include figure files
\usepackage{subfig} 
\usepackage{dcolumn}% Align table columns on decimal point 
\usepackage{bm}% bold math
\usepackage{latexsym}
\usepackage{float}

\def\av#1{\langle#1\rangle}

\begin{document} 

\title[Phase transition in rich-get-richer mechanism]{Phase transition in the rich-get-richer mechanism due to finite-size effects}  

\author{James P Bagrow$^1$, Jie Sun$^2$ and 
Daniel ben-Avraham$^1$} 
\address{$^1$ Department of Physics, Clarkson University,
Potsdam, NY 13699-5820, USA} 
\address{$^2$ Department of Mathematics \& Computer Science,  Clarkson University, Potsdam, NY 13699-5815, USA} 
\eads{\mailto{bagrowjp@clarkson.edu}, \mailto{sunj@clarkson.edu}, \mailto{qd00@clarkson.edu}}

\begin{abstract} 
The rich-get-richer mechanism (agents increase their``wealth" randomly at a rate proportional to their holdings) is often invoked to explain the Pareto power-law distribution observed in many physical situations, such as the degree distribution of growing scale free nets.   We use two different analytical approaches, as well as numerical simulations, to study the case where the number of agents is fixed and finite (but large), and the rich-get-richer mechanism is invoked a fraction $r$ of the time (the remainder of the time wealth is disbursed by a homogeneous process).   At short times, we recover the Pareto law observed for an unbounded number of agents.  In later times, the (moving) distribution can be scaled to reveal a phase transition with a Gaussian asymptotic form for $r<\case{1}{2}$, and a Pareto-like tail (on the positive side) and a novel stretched exponential decay (on the negative side) for $r>\case{1}{2}$. 
\end{abstract} 

\pacs{
02.50.Ey,	%Stochastic processes
05.70.Fh,	%Phase transitions: general studies
89.75.Da	%Systems obeying scaling laws
}

\submitto{\JPA} 
\maketitle

\section{Introduction}
The rich-get-richer mechanism, or preferential attachment, is often invoked to explain the ubiquity of Pareto distributions in complex networks and in natural and man-made phenomena~\cite{b-a,dorogovtsev,newman}.  For example, in the Barab\'asi-Albert model for complex networks~\cite{bamodel} a power-law tail in the degree distribution of the nodes, $P(k)\sim k^{-3}$,
arises when new nodes are attached to the net, one at a time, to one of the existing nodes with a probability
proportional to their degree.  Crucial to the emergence of this power-law tail, however, is also the fact
that the net {\it grows\/} with time (new nodes are attached) --- preferential attachment alone is not the only
necessary ingredient.

In this paper we study the consequences of preferential attachment without growth, that is, when ``wealth" is disbursed to a {\it finite\/} group of $A$ agents by the rich-get-richer mechanism.  More precisely, wealth is incremented by the rich-get-richer mechanism only a fraction $r$ of the time; the remainder of the time
(a fraction $1-r$) wealth is disbursed homogeneously at random.  For an infinite number of agents, $1-r$ is simply the rate at which new agents are introduced (are given their first unit of wealth), and there results a Pareto distribution of wealth, $P(k)\sim k^{-\lambda}$, with
$\lambda = 1+1/r$.  When $A$ is finite, however, the homogeneous process may allot wealth to agents that had already been introduced.  The case when $r=0$ and wealth is disbursed only by the homogeneous process results in a Gaussian distribution of wealth --- quite different from the power-law
distribution found in the limit of $A\to\infty$.  The two extremes, of $A\to\infty$, and $A<\infty$ and $r=0$, dominate the general case, pulling the system toward opposite poles and resulting in a kinetic phase transition: the long-time asymptotic distribution of wealth is Gaussian for $r<\case{1}{2}$, but has a power-law tail for $r>\case{1}{2}$.  In all cases the wealth distribution can be written in scaling form, and the shape
of the scaling function is characterized by various $r$-dependent exponents that we obtain analytically.

The rest of this paper is organized as follows.  In \sref{model} we present the model and analyze the two
special limits that color its behavior.  This section serves also to illustrate the two mathematical techniques that we employ for the model's analysis: the {\it master equation\/}  (for the number of agents with a given wealth) and the {\it rate equation\/} approach (called also  {\it mean-field} by other researchers).  In the latter
approach one uses a rate equation to compute the expected wealth of a particular agent as a function of the time of its introduction. The wealth distribution may then be obtained from the distribution of the introduction times.  The general case is treated in \sref{general}.  We derive several analytical results and compare to computer simulations, as well as discuss the range of validity of the rate equation approach.  We conclude with a summary of our findings and a discussion in \sref{discussion}.

\section{The model}\label{model}

We analyze the following toy model of wealth allotment.  A single wealth unit is given per unit time to a random member out of $A$ agents.  With probability $r$ the agent is chosen by the {\it  rich-get-richer\/} mechanism, proportional to its current wealth; with probability $1-r$ the agent is chosen {\it uniformly at random\/}.  Thus $r$ is a tunable parameter controlling the relative strengths of these two processes.  We assume that initially each agent has zero wealth.  Let us start by exploring the two limits  of $A \to \infty$ and $r\to0$.  These exactly solvable extremes yield markedly different behavior and underlie the essential physics of the system when $A$ and $r$ are finite: for $r<\case{1}{2}$ the uniform random process dominates, while for $r>\case{1}{2}$ the rich-get-richer process gets the 
upper hand.

\subsection{The limit of $A\to\infty$}\label{Ainf}

For $A \to \infty$, agents with positive wealth form a set of measure zero for any finite time.  Thus choosing an agent uniformly at random is equivalent to the ``birth'' of a new agent~$i$ with wealth $k_{i}=1$.  We analyze this case with a master equation approach, following the techniques and notation of~\cite{kr}.  The number of agents with wealth $k>0$ at time $t$, $N_{k}(t)$, obeys the master equation:
\begin{equation}
	\frac{d}{dt} N_k =  (1-r)\delta_{k,1} +  \frac{r}{\sum_{k'} k'N_{k'} } \Big[(k-1)N_{k-1} - k N_k \Big] \,.
	\label{eqn:InfRate}
\end{equation}
Since one wealth unit is disbursed per unit time, $\sum_{k} k N_{k} = t$.  This limit is characterized by both growth and preferential attachment,  hence we expect a power-law distribution of wealth.  This master equation has been studied 
before~\cite{kr} and indeed it is easy to show that the actual distribution of wealth is 
\begin{equation}
P(k) = N_k (t) / N(t) \sim k^{-\lambda};  \qquad \lambda = 1 + \frac{1}{r}\, .
\end{equation}

One can also derive this distribution using a rate equation for the average wealth of agent $i$:
\begin{equation}
	\label{ki_inf_rate}
	\frac{d}{d t}k_{i}(t)=\frac{r}{t}k_{i}(t)\,,
\end{equation}
since $\sum_j k_j = t$.  Because $A\to\infty$, the rate at which agent $i$ is selected by the homogeneous random process
is zero, while the rate of selection by the rich-get-richer mechanism is $k_i/\sum_j k_j=k_i/t$.  Following the analysis in~\cite{baj}, one can show that
\begin{equation}
	P(k)= \frac{1}{r} k^{-1-1/r},\qquad k\geq1\,,
\end{equation}
i.e., a Pareto distribution with the same power-law tail.  Note that this distribution is properly normalized (taking $k$ to be a continuous variable) and that its first moment agrees with that from the master equation. 

While the results for both the master equation and the rate equation approach are equivalent for $A \to \infty$, we will show that this is no longer true for $A <\infty$. 

%Assuming that
%agent $i$ got his first unit of wealth at time $t_i$, the initial condition for~\eref{ki_inf_rate} is 
%$k_{i}(t_{i})=1$.  Thus, 
%\begin{equation}
%\label{ki(t)}
%k_i(t_i)=\left(\frac{t}{t_i}\right)^r,
%\end{equation}
%is a monotonous decreasing function of $t_i$.  It follows that the probability that $k_i>k$ is the same as the probability that $t_i<T$, where $k_i(t_i=T)=k$.
%In other words,
%\begin{equation}
%\chi(k)\equiv {\rm Pr}(k_i>k)=\int_k^{\infty} P(k')\,dk'={\rm Pr}(t_i<T)\,.
%\end{equation}
%But $T=tk^{-1/r}$, from \eref{ki(t)}, and since the probability that agent $i$ gets its first unit of wealth (by the
%homogeneous random process) is uniform in time, ${\rm Pr}(t_i<T)=T/t=k^{-1/r}$.  We then have

\subsection{The limit of $r\to0$}

When the number of agents $A$ is finite, the $N_k$ obey the normalization condition
\begin{equation}
\sum_{k=0}^{\infty} N_k(t)=A\,,
\end{equation}
where now we include in the counting agents with zero wealth ($k=0$),
and the distribution of wealth is $P(k,t)=N_k(t)/A$.  The mean wealth per agent increases linearly with time:
\begin{equation}
\av{k}=\frac{t}{A}\,.
\end{equation}
Consider the limit of $r\to0$, where wealth is disbursed only by the homogeneous random process.
The corresponding master equation is
\begin{equation}
\frac{d}{dt} N_k = \frac{1}{A}\Big( N_{k-1} - N_k\Big)\,,
\label{MasterEqr0}
\end{equation}
with initial and boundary conditions $N_k(0)=A\delta_{k,0}$  and $N_{-1}(t)=0$.
This is a simple Poisson process, as confirmed by the solution of \eref{MasterEqr0}:
\begin{equation}
N_k(t)=A\frac{(t/A)^k}{k!}\e^{-t/A}\,.
\end{equation}
For $t\gg A$ we apply the Sterling approximation to obtain the distribution
\begin{equation}
\label{Pr=0}
P(k,t)=\frac{1}{\sqrt{2\pi (t/A)}}\,\exp\!\left(-\frac{1}{2}A\frac{(k-t/A)^2}{t}\right)\,.
\end{equation}
Thus, $P(k,t)$ has a power-law tail in the limit $A\to\infty$ (\sref{Ainf}), but is Gaussian when $A$ is 
finite and $r\to0$.

\section{Finite $A$ and $r$}\label{general}
\subsection{Master equation approach}

In the general case of $A<\infty$ and $r>0$ the master equation for the process is
\begin{equation}
\frac{d}{dt} N_k = \frac{1-r}{A}\Big( N_{k-1} - N_k\Big)+ \frac{r}{t}\Big((k-1)N_{k-1} - k N_k \Big)\,.
\label{MasterGen}
\end{equation}
The system is then simultaneously pulled toward the two different limiting behaviors analyzed in 
\sref{model}.  We will show that for $r>\case{1}{2}$ the rich-get-richer mechanism dominates the process and the
wealth distribution develops a power-law tail (as for the limit of $A\to\infty$), while for $r<\case{1}{2}$ the 
homogeneous random process dominates and the wealth distribution tends to a Gaussian (as for $r\to0$).  Because 
$A$ is finite, $\av{k}=t/A$ increases linearly with time.  The width of the distribution of $k$ around the average grows like $t^{\alpha}$, where the scaling exponent  $\alpha=r$ for $r>\case{1}{2}$ and $\alpha=\case{1}{2}$ for $r<\case{1}{2}$.  At the transition point, $r=\case{1}{2}$, the width scales as $\sqrt{t\ln t}$.

To see this, begin by approximating the discrete distribution $N_{k}(t)$ by its continuous counterpart, 
 $P(k,t)$.  Expanding to first-order, equation \eref{MasterGen} now reads
\begin{equation}
	\frac{\partial}{\partial t} P(k,t) = -\frac{1-r}{A}\frac{\partial}{\partial k}P- \frac{r}{t}\frac{\partial}{\partial k}(kP)\,,
\end{equation}
and the method of characteristics yields the scaling solution
\begin{equation}
	P(k,t)=t^{-\alpha}f\left(\frac{k-t/A}{t^{\alpha}}\right);\qquad\alpha=r \,.
	\label{scaling_form}
\end{equation} 
This, however, cannot be true for all values of $r$, as it disagrees with the distribution \eref{Pr=0} found for $r=0$, where the scaling exponent is $\alpha=\case{1}{2}$ instead of $\alpha=r=0$.  The reason for this discrepancy
is that in this case the Kramers-Moyal expansion~\cite{vk} of \eref{MasterGen} must be carried out beyond the first order.  Indeed, on substituting the scaling form \eref{scaling_form} in the master equation (with unspecified $\alpha$), and carrying out the expansion to second-order, we find
\begin{equation}
\label{k-m}
(\alpha-r)t^{2\alpha}f(x)+(\alpha-r)t^{2\alpha}xf'(x)+\frac{1}{2A}tf''(x)=0\,,
\end{equation}
where  prime denotes differentiation with respect to $x=(k-t/A)/t^{\alpha}$, and we have omitted terms
proportional to $t^{\alpha}$ (these are negligible compared to $t^{2\alpha}$, as $t\to\infty$).
If $\alpha>\case{1}{2}$, the term proportional to $t$ can be neglected in the long-time limit, and \eref{k-m} is satisfied provided that $\alpha=r$.  Thus, the scaling form \eref{scaling_form} is valid only for $r>\case{1}{2}$.  For $r<\case{1}{2}$, however, the second-order term in \eref{k-m} may not be ignored.  The only non-trivial way to cancel out the time dependence is then to have $t^{2\alpha}=t$.  Thus, for $r<\case{1}{2}$ the scaling exponent is $\alpha=\case{1}{2}$.
At the transition point, $r=\case{1}{2}$, there is no way to get rid of the time dependence in \eref{k-m} with the scaling form \eref{scaling_form}.  Taking a cue from other phase transitions we guess a scaling form with logarithmic dependence:
\begin{equation}
P(k,t)=\frac{1}{(t\ln t)^{\alpha}}f\left(\frac{k-t/A}{(t\ln t)^{\alpha}}\right);\qquad r=\frac{1}{2} \,.
\label{scaling_half}
\end{equation} 
On expanding the master equation with this scaling form the leading behavior in time cancels out,
provided that the scaling exponent is $\alpha=\case{1}{2}$.  The next largest terms (smaller by a $\ln t$ factor), yield the equation
\begin{equation}
\label{k-m1/2}
f(x)+xf'(x)+\frac{1}{A}f''(x)=0\,;\qquad r=\frac{1}{2} \,,
\end{equation}
where now $x=(k-t/A)/\sqrt{t\ln t}$.
In all three cases (for $r$), expanding to third- or higher-order yields additional subdominant terms.
From the largest subdominant term  one can deduce how fast the system reaches the scaling
regime:  the transient dies off as $t^{-(2r-1)}$ for $r>\case{1}{2}$, as $t^{-1/2}$ for $r<\case{1}{2}$, and as $(\ln t)^{-1}$
for $r=\case{1}{2}$.  Thus at the transition point, $r=\case{1}{2}$, there occurs a {\it critical slowing down\/}
as the system creeps into the eventual scaling regime logarithmically slow. 

For $r<\case{1}{2}$ we can use \eref{k-m} to find out $f(x)$ and show that the limiting form of the wealth distribution
is Gaussian:
\begin{equation}
\label{gauss}
P(k,t)\to\sqrt{\frac{A(1-2r)}{2\pi t}}\,\exp\!\left[-\frac{1}{2}A(1-2r)\frac{(k-t/A)^2}{t}\right]\,;\qquad r<\frac{1}{2}\,,
\end{equation}
as $t\to\infty$.  The divergence of the width of this distribution as $r\to\case{1}{2}$ is reconciled with the fact that
at the limit $r=\case{1}{2}$ the scaling parameter picks up a (diverging) logarithmic component.
The scaling function is still Gaussian, as can be deduced from~\eref{k-m1/2}:
\begin{equation}
\label{Phalf}
P(k,t)\to\sqrt{\frac{A}{2\pi \,t\ln t}}\,\exp\!\left[-\frac{1}{2}A\frac{(k-t/A)^2}{t\ln t}\right]\,;\qquad r=\frac{1}{2}\,.
\end{equation}
For $r>\case{1}{2}$ equation \eref{k-m} yields a tautology and one is unable to determine $f(x)$.  It is possible,
nevertheless, to infer the limiting behavior:
\begin{equation}
\label{asympf}
f(x)\sim
\cases{
x^{-1-1/r} & $x\to\infty$,\cr
\exp\left[-(1-r)(A^r|x|)^{\frac{1}{1-r}}\right] & $x\to -\infty$,
}\quad r>\frac{1}{2}\,.
\end{equation}
The limit for $x\to\infty$ follows from comparing the distribution $P(k,t)$ for the case of $A\to\infty$ with 
$f(x)|_{A\to\infty}=f(kt^{-r})$.
For $x\to-\infty$, we observe that the density of agents with zero wealth decays as 
$N_0\sim\exp[-(1-r)t/A]$, see equation \eref{psi},
and we compare to $f(x)|_{k=0}=f(-t^{1-r}/A)$, leading to the second line of \eref{asympf}.
An alternative derivation is presented next, using the rate equation approach.

\subsection{Rate equation approach}

The rate equation for the wealth of agent $i$, in the general case, is
\begin{equation}
\frac{d}{d t}k_{i}(t)=\frac{1-r}{A}+\frac{r}{t}k_{i}(t)\,,
\end{equation}
with initial condition $k_{i}(t_{i})=1$.  The solution,
\begin{equation}
\label{ki}
k_i(t_i)=\Big(1-\frac{t_i}{A}\Big)\left(\frac{t}{t_i}\right)^r+\frac{t}{A}\,,
\end{equation}
is monotonously decreasing in $t_i$.

The probability  $\psi(t)$ that an agent has still zero wealth at time $t$ satisfies the equation
\begin{equation}
\label{psi}
\frac{d}{dt}\psi(t)=-\frac{1-r}{A}\psi(t)\,,
\end{equation}
so $\psi(t)=\exp[-(1-r)t/A]$.  It follows that the probability that agent $i$ has been introduced (gets its first unit of wealth) by time 
$T$, given that it has been introduced by time $t$, is
\begin{equation}
\chi(T)=\frac{1-e^{-\frac{1-r}{A}T}}{1-e^{-\frac{1-r}{A}t}}\,.
\end{equation}
Note that this has the limit $T/t$, as $A\to\infty$, that is normally used for this case~\cite{baj}.

Finally,
$P(k,t)=-\partial \chi(T)/\partial k$, where $T(k)$ is the solution to $k_i(T)=k$.  Since~\eref{ki} cannot
be inverted analytically (other than for special values of $r$), we express $P$ in parametric form:
$P(k(T),t)=-\partial \chi(T)/\partial k=-(dk_i/dt_i|_{t_i=T})^{-1}\partial \chi(T)/\partial T$, and $k(T)$
is obtained by putting $t_i=T$ in \eref{ki}. The 
wealth distribution in parametric form is then
\begin{equation}
\label{param}
x(T)=\Big(1-\frac{T}{A}\Big)T^{-r}\,,\qquad
f(T)=\frac{(1-r)T^{1+r}}{Ar+(1-r)T}e^{-\frac{1-r}{A}T}\,,
\end{equation}
where we have used the scaled expressions $x=(k-t/A)/t^r$ and $f=t^rP$, taking the limit of $t\to\infty$ at the end (the fact that the limit exists and is finite confirms this scaling).  

It is now easy to verify the asymptotic behavior~\eref{asympf}.  The limit $x\to\infty$ corresponds to $T\to0$.
In this limit, the second equation of~\eref{param} gives $f\sim T^{1+r}$.  But since $T\sim x^{-1/r}$, from
the first equation, we conclude that $f\sim x^{-1-1/r}$.  The limit $x\to-\infty$ corresponds 
to $T\to\infty$. In this limit, the second equation of~\eref{param} gives $f\sim\exp[-(1-r)T/A]$,
while from the first equation $x\sim -(1/A)T^{1-r}$.  We conclude that $f\sim\exp[-(1-r)(A^r|x|)^{1/(1-r)}]$.

\begin{figure}[t]
   \centering
  \subfloat{\label{fig:gull}%
  { % GNUPLOT: LaTeX picture with Postscript
\begingroup%
\makeatletter%
\newcommand{\GNUPLOTspecial}{%
  \@sanitize\catcode`\%=14\relax\special}%
\setlength{\unitlength}{0.0500bp}%
\begin{picture}(3600,2520)(0,0)%
  \includegraphics{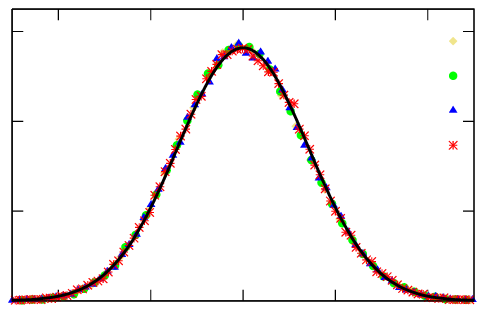}
  \put(3020,1496){\makebox(0,0)[r]{\strut{}$8 \cdot 10^5$}}%
  \put(3020,1696){\makebox(0,0)[r]{\strut{}$4 \cdot 10^5$}}%
  \put(3020,1896){\makebox(0,0)[r]{\strut{}$2 \cdot 10^5$}}%
  \put(3020,2096){\makebox(0,0)[r]{\strut{}$1 \cdot 10^5$}}%
  \put(1930,100){\makebox(0,0){\strut{}$(k - t/A) / t^{1/2}$}}%
  \put(200,1440){%
%
  \makebox(0,0){\strut{}$t^{1/2} P(x)$}%
%
  }%
  \put(2994,400){\makebox(0,0){\strut{} 0.4}}%
  \put(2462,400){\makebox(0,0){\strut{} 0.2}}%
  \put(1930,400){\makebox(0,0){\strut{} 0}}%
  \put(1398,400){\makebox(0,0){\strut{}-0.2}}%
  \put(866,400){\makebox(0,0){\strut{}-0.4}}%
  \put(480,2151){\makebox(0,0)[r]{\strut{} 3}}%
  \put(480,1634){\makebox(0,0)[r]{\strut{} 2}}%
  \put(480,1117){\makebox(0,0)[r]{\strut{} 1}}%
  \put(480,600){\makebox(0,0)[r]{\strut{} 0}}%
\end{picture}%
\endgroup
}}
  \subfloat{\label{fig:tiger}%
   {% GNUPLOT: LaTeX picture with Postscript
\begingroup%
\makeatletter%
\newcommand{\GNUPLOTspecial}{%
  \@sanitize\catcode`\%=14\relax\special}%
\setlength{\unitlength}{0.0500bp}%
\begin{picture}(3600,2520)(0,0)%
  \includegraphics{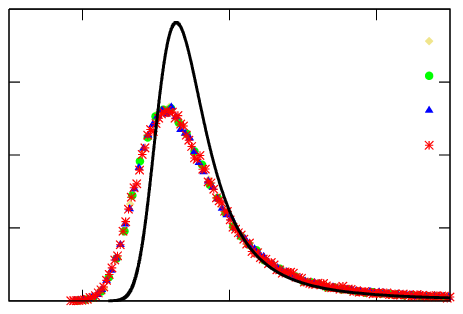}
  \put(3020,1496){\makebox(0,0)[r]{\strut{}$8\cdot 10^5$}}%
  \put(3020,1696){\makebox(0,0)[r]{\strut{}$4\cdot 10^5$}}%
  \put(3020,1896){\makebox(0,0)[r]{\strut{}$2\cdot 10^5$}}%
  \put(3020,2096){\makebox(0,0)[r]{\strut{}$1\cdot 10^5$}}%
  \put(1990,100){\makebox(0,0){\strut{}$(k - t/A) / t^{r}$}}%
  \put(200,1440){%
%
  \makebox(0,0){\strut{}$t^{r} P(x)$}%
%
  }%
  \put(2837,400){\makebox(0,0){\strut{} 0.1}}%
  \put(1990,400){\makebox(0,0){\strut{} 0}}%
  \put(1143,400){\makebox(0,0){\strut{}-0.1}}%
  \put(600,2280){\makebox(0,0)[r]{\strut{} 20}}%
  \put(600,1860){\makebox(0,0)[r]{\strut{} 15}}%
  \put(600,1440){\makebox(0,0)[r]{\strut{} 10}}%
  \put(600,1020){\makebox(0,0)[r]{\strut{} 5}}%
  \put(600,600){\makebox(0,0)[r]{\strut{} 0}}%
\end{picture}%
\endgroup
}}\\
  \subfloat{\label{fig:mouse}%
 { % GNUPLOT: LaTeX picture with Postscript
\begingroup%
\makeatletter%
\newcommand{\GNUPLOTspecial}{%
  \@sanitize\catcode`\%=14\relax\special}%
\setlength{\unitlength}{0.0500bp}%
\begin{picture}(7200,5040)(0,0)%
  \includegraphics{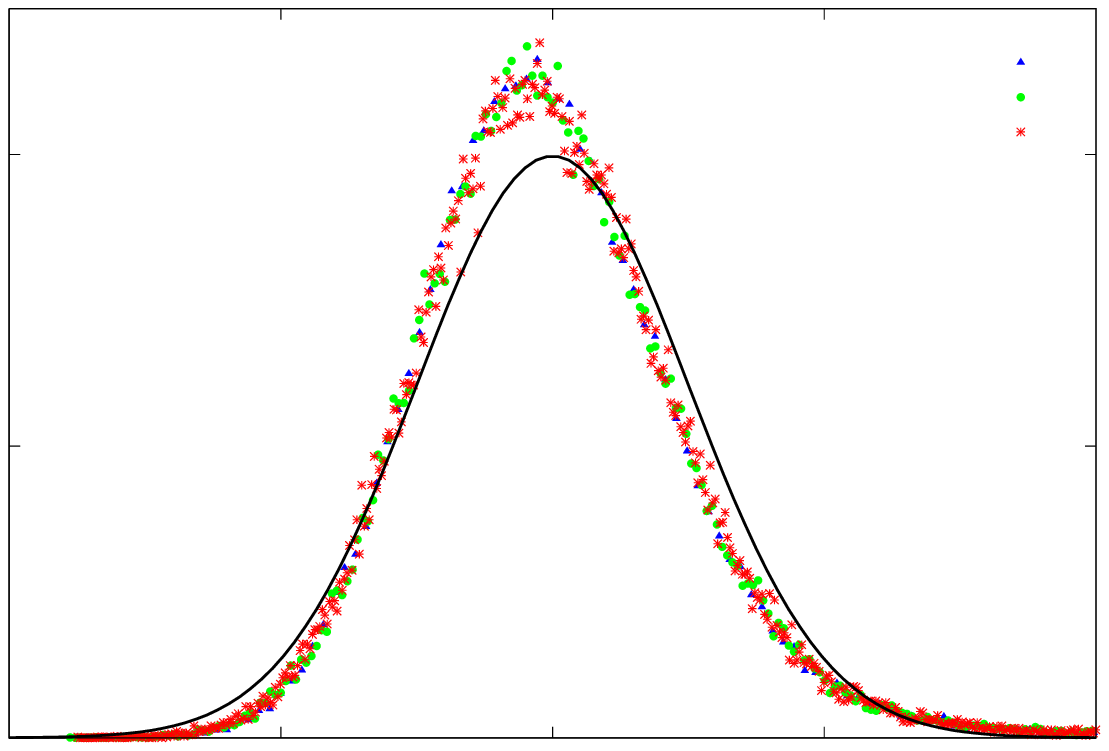}
  \put(6307,4090){\makebox(0,0)[r]{\strut{}$16\cdot 10^7$}}%
  \put(6307,4290){\makebox(0,0)[r]{\strut{}$4\cdot 10^7$}}%
  \put(6307,4490){\makebox(0,0)[r]{\strut{}$1\cdot 10^7$}}%
  \put(3730,100){\makebox(0,0){\strut{}$(k - t/A) / (t \ln t)^{1/2}$}}%
  \put(200,2700){%
%
  \makebox(0,0){\strut{}$(t \ln t)^{1/2} P(x)$}%
%
  }%
  \put(6860,400){\makebox(0,0){\strut{} 0.4}}%
  \put(5295,400){\makebox(0,0){\strut{} 0.2}}%
  \put(3730,400){\makebox(0,0){\strut{} 0}}%
  \put(2165,400){\makebox(0,0){\strut{}-0.2}}%
  \put(600,400){\makebox(0,0){\strut{}-0.4}}%
  \put(480,3960){\makebox(0,0)[r]{\strut{} 4}}%
  \put(480,2280){\makebox(0,0)[r]{\strut{} 2}}%
  \put(480,600){\makebox(0,0)[r]{\strut{} 0}}%
\end{picture}%
\endgroup
}}
   \caption{(color online) Scaling of the wealth distribution in each of the two phases at $r=\case{1}{4}$ (top left), $r=\case{3}{4}$ (top right), and at the transition point $r=\case{1}{2}$ (bottom).  Convergence to the scaling forms is rapid for $r=\case{1}{4}$ and $r=\case{3}{4}$ but logarithmically slow for $r=\case{1}{2}$ --- note that in the latter case the data (for exponentially increasing times) is slowly creeping toward the Gaussian limit of \eref{Phalf} (solid line).   The theoretical limit of \eref{gauss} (solid line) fits the case of $r=\case{1}{4}$ perfectly, but the prediction \eref{param} from the rate equation approach (solid line) fits the case of $r=\case{3}{4}$ only qualitatively (besides agreeing with the overall scaling).}
   \label{scaling}
\end{figure}

Clearly, the foregoing rate equation method does not apply to $0\leq r \leq\case{1}{2}$, for it fails to reproduce the appropriate scaling forms in this range.  Thus the rate equation approach is viable only when the second-order in the Kramers-Moyal expansion of the corresponding master equation may be neglected.  In \fref{scaling} we show numerical simulations for $r$ below, above, and at the transition point.  The results confirm the scaling forms found
analytically above.  For $r<\case{1}{2}$ convergence to the Gaussian pdf is relatively fast, while the
critical slowing down at the transition point, $r=\case{1}{2}$, prevents us from attaining the analytical limit~\eref{Phalf} in practice. For $r>\case{1}{2}$ convergence to the scaling pdf is again quick, however the explicit form predicted by the rate equation approach is correct only qualitatively: we ascribe this to the fact that the second-order is implicitly missing in this approach.

\section{Summary and discussion}\label{discussion}

In summary, we have studied a model of  a finite number $A$ of agents that accrue ``wealth" by the rich-get-richer mechanism a fraction $r$ of the time (wealth is disbursed homogeneously randomly the remainder $1-r$ of the time).  In the early time regime, or, equivalently, when $A\to\infty$ there results a Pareto distribution of the wealth $k$: $P(k)\sim k^{-\lambda}$, with $\lambda=1+1/r$.  In the long time asymptotic limit, the system is attracted to one of two opposite poles, and there is a kinetic phase transition as a function of the 
parameter $r$.  If $r<\case{1}{2}$, the distribution tends to a Gaussian of width $t/[(1-2r)A]$.  If $r>\case{1}{2}$, the distribution keeps its power-law tail $\sim k^{-1-1/r}$ for large~$k$.

In all cases the wealth distribution tends to an asymptotic scaling form as a function of $x=(k-\av{k})/w(t)$,
where $\av{k}=t/A$ is the average wealth amassed by an agent by the time $t$, and 
$w(t)=t^{\alpha}$ is a measure of the width of the distribution.  The exponent $\alpha$ undergoes a phase transition: $\alpha=\case{1}{2}$ for $r<\case{1}{2}$, and $\alpha=r$ for $r>\case{1}{2}$.  At the transition
point, $r=\case{1}{2}$, there appear logarithmic corrections: $w(t)=(t\ln t)^{1/2}$.

The scaling form of the wealth distribution $f(x)=t^rP$ in the regime $r>\case{1}{2}$ is characterized by two more exponents (in addition to the width exponent $\alpha=r$):
$f(x)\sim x^{-1-1/r}$ for $x\to\infty$, and $f(x)$ decays as a stretched-exponential, with power $1/(1-r)$, as
$x\to-\infty$.
Finally, the approach  to the eventual scaling form $\sim t^{-z}$ is characterized by a fourth exponent:
$z=\case{1}{2}$ for $r<\case{1}{2}$, and $z=2r-1$ for $r>\case{1}{2}$.  At the transition point convergence to the scaling form proceeds exceedingly slow, $\sim1/\ln t$, in a fashion reminiscent of critical slowing down
in equilibrium phase transitions.

Several applications come to mind.  For example, complex networks could be grown according to this model
where the nodes are fixed at the outset (corresponding to the $A$ agents) and links are connected to the
nodes by a proper mix of homogeneous selection and preferential attachment.  For $r>\case{1}{2}$ one could thus create scale-free nets with a fixed degree distribution exponent and a fixed number of nodes, and with a tunable average connectivity $\av{k}=t/A$ that grows linearly with time.  Wealth distributions with a stretched-exponential decay on one side and a power-law decay on the other, such as we find for $r>\case{1}{2}$, are regularly observed in various economic settings~\cite{cc,ccc}.

An intriguing finding concerns the method of rate equations, or mean-field method that is often used to obtain the degree distribution of complex networks~\cite{baj,jkk,zrc,crb,zetal}.  Our analysis suggests that this method is only valid when the second-order terms in the Kramers-Moyal expansion of the master equation for the system may be safely neglected.  Even then the method yields results that scale correctly but that are otherwise only qualitatively correct, at least in our case.  Perhaps the most important open problem is to establish the range of validity of the rate equation approach more rigorously, and to find ways to extend it to the cases where it fails.

\ack
We thank E.~Bollt,  H.~Rozenfeld and J.~Skufca for discussions.  We gratefully acknowledge the NSF for partial funding (DbA)
and for their support of JPB as a NSF Graduate Research Fellow.

\end{document}